\documentclass[aps,prl,showpacs,twocolumn,a4paper]{revtex4}
\usepackage{graphicx}
\usepackage{amsmath}
\usepackage{amsfonts}

\begin{document}

\title{Topological-Fermi-Liquid to Quantum-Hall-Liquid Transitions: $p$-Band and $d$-Band Fermions in a Magnetic Field}
\author{Yi-Fei Wang$^{1}$ and Chang-De Gong$^{1,2}$}
\affiliation{$^1$Center for Statistical and Theoretical Condensed
Matter Physics, and Department of Physics, Zhejiang Normal
University, Jinhua 321004, China\\$^2$National Laboratory of Solid
State Microstructures and Department of Physics, Nanjing University,
Nanjing 210093, China}
\date{\today}

\begin{abstract}
We find that in a multi-orbital system with intraorbital and
interorbital hopping integrals, the Hall conductance exhibits
various topological quantum phase transitions (QPTs) induced by
on-site orbital polarization: integer quantum Hall (IQH) plateau
transitions, and topological Fermi liquid to IQH transitions. Such
topological QPTs are demonstrated in two systems: a $p$-band
spinless fermionic system realizable with ultracold atoms in optical
lattice, and a $d$-band spinful fermionic system closely related to
giant orbital Hall effects in transition metals and their compounds.

\end{abstract}

 \maketitle

{\it Introduction.---} Orbital is an additional degree of freedom
independent of charge and spin, and is characterized by orbital
degeneracy and spatial anisotropy. Recently, it is proposed that
novel $p$-orbital physics exist in optical lattices, and earlier
attention has been paid on bosons in the first excited $p$-orbital
bands~\cite{Isacsson}. Rapid experimental advances in loading and
controlling alkali atoms on the excited bands makes the $p$-orbital
physics truly fascinating~\cite{Kohl,Muller,Anderlini}, and
especially a metastable $p$-orbital bosonic system has been
realized~\cite{Muller}. Correlated fermions in the $p$-orbital bands
possess more arresting behaviors, including Wigner crystallization,
orbital ordering and frustration~\cite{Wu2,Zhao,Wu3}. Even for
non-interacting $p$-band spinless fermions, non-trivial topological
band structures arises from lifting the orbital degeneracy, and
Haldane's quantum Hall model without Landau levels~\cite{Haldane1}
can be realized~\cite{Wu4}.

In parallel, for transition metal oxides, the relevant active
orbitals are the partially filled five $d$-orbitals. Recently, giant
spin Hall and orbital Hall effects have been found in transition
metals and their compounds~\cite{Valenzuela,Stern,Kimura}, and many
theoretical studies are based on realistic multiband $4d$ or $5d$
models\cite{Kontani,Tanaka,Guo}. A possible intrinsic origin of
these giant Hall effects is the ``orbital Aharonov-Bohm phase''
which is induced by the on-site spin-orbital coupling (SOC) and the
phase of interorbital hopping integrals characteristic of
$d$-orbital systems, and therefore the conduction electrons are
subject to an effective spin-dependent magnetic
field~\cite{Kontani,Tanaka}.

Given the intensive current interest in possible novel $p$-orbital and
$d$-orbital physics which has not appeared in single-orbital systems,
we are motivated to study magneto-transport
properties of multi-orbital systems with intraorbital and
interorbital hopping integrals modulated by an external magnetic
flux, and find that the Hall conductance (HC) may exhibit various
topological quantum phase transitions (QPTs) induced by on-site
orbital polarization: integer quantum Hall (IQH) plateau
transitions, and topological Fermi liquid (TFL) to IQH transitions.
An IQH effect, characterized by a nonzero Chern
integer~\cite{Thouless},
only occurs when the Fermi level lies in an energy gap, and is a
Fermi sea property. While the non-quantized part of the HC of a TFL
is characterized by a Berry phase accumulated by adiabatic motion of
quasiparticles on the Fermi surface (FS), and is
thus purely a Fermi liquid property
~\cite{Haldane2}. Such topological QPTs are demonstrated in two
systems: a $p$-band spinless fermionic system which is proposed to
be realized with ultracold atoms in optical lattice~\cite{Wu4}, and
a $d$-band spinful fermionic system which is closely related to
giant orbital Hall effect in transition metals and their
compounds~\cite{Kontani,Tanaka}.

{\it Formulation.---}The first model is a $p$-band system of $p_{x}$
and $p_{y}$ orbitals in a 2D square lattice filled with spinless
fermions coupled with an artificial uniform magnetic
flux~\cite{Wu2,Zhao,Wu3,Wu4,Umucalilar}:
\begin{eqnarray}
H_{p}&=&\sum_{\mathbf{r}}\sum_{\mu,\nu=x,y}\left[t_{\parallel}\delta_{\mu\nu}
-t_{\perp}\left(1-\delta_{\mu\nu}\right)\right]\nonumber\\
&&\times\left[p^{\dagger}_{\mu,\mathbf{r}}p_{\mu,\mathbf{r}+\mathbf{e}_{\nu}}\exp\left(i\phi_{\mathbf{r},\mathbf{r}+\mathbf{e}_{\nu}}\right)
+\textrm{H.c.}\right]\nonumber\\
&&+\lambda\sum_{\mathbf{r}}\left(ip^{\dagger}_{x,\mathbf{r}}p_{y,\mathbf{r}}-ip^{\dagger}_{y,\mathbf{r}}p_{x,\mathbf{r}}\right),
\label{e.1}
\end{eqnarray}
where $p^{\dagger}_{\mu,\mathbf{r}}$ is a fermion creation
operator of $\mu=x,y$ $p$-orbital at site $\mathbf{r}$. 
$t_{\parallel}$ and $t_{\perp}$ are the nearest neighbor (NN)
hopping integrals in the longitudinal and transverse directions,
respectively, to each $p$-orbital orientation.
$t_{\parallel},t_{\perp}>0$, and $t_{\perp}$ is conventionally one
order of magnitude smaller than $t_{\parallel}$. In the following,
$t_{\parallel}$ will be taken as the unit of energy. A finite
$\lambda$ induces the rotation of each site around its own center,
thus gives rise to orbital polarization by lifting the degeneracy
between $p_x\pm ip_y$ orbitals~\cite{Wu4,Umucalilar}.

The second model is a $d$-band system of $d_{xz,\sigma}$ and
$d_{yz,\sigma}$ orbitals (the notation is simplified as
$d_{xz,\sigma}\equiv d_{x\sigma}$ and $d_{yz,\sigma}\equiv
d_{y\sigma}$) in a 2D square lattice filled with spinful fermions
coupled with a uniform magnetic flux~\cite{Kontani,Tanaka}:
\begin{eqnarray}
H_{d}&=&-t_{\parallel}\sum_{\mathbf{r},\sigma,\mu}
\left[d^{\dagger}_{\mu\sigma,\mathbf{r}}d_{\mu\sigma,\mathbf{r}+\mathbf{e}_{\mu}}\exp\left(i\phi_{\mathbf{r},\mathbf{r}+\mathbf{e}_{\mu}}\right)+\textrm{H.c.}\right]\nonumber\\
&+&t^{\prime}\sum_{\mathbf{r},\sigma}
\left[d^{\dagger}_{x\sigma,\mathbf{r}}d_{y\sigma,\mathbf{r}\pm\mathbf{e}_{x}\pm\mathbf{e}_{y}}\exp\left({i\phi_{\mathbf{r},\mathbf{r}\pm\mathbf{e}_{x}\pm\mathbf{e}_{y}}}\right)+\textrm{H.c.}\right]\nonumber\\
&-&t^{\prime}\sum_{\mathbf{r},\sigma}
\left[d^{\dagger}_{x\sigma,\mathbf{r}}d_{y\sigma,\mathbf{r}\pm\mathbf{e}_{x}\mp\mathbf{e}_{y}}\exp\left({i\phi_{\mathbf{r},\mathbf{r}\pm\mathbf{e}_{x}\mp\mathbf{e}_{y}}}\right)+\textrm{H.c.}\right]\nonumber\\
&+&\lambda\sum_{\mathbf{r}}\left(id^{\dagger}_{x\downarrow,\mathbf{r}}d_{y\downarrow,\mathbf{r}}-id^{\dagger}_{x\uparrow,\mathbf{r}}d_{y\uparrow,\mathbf{r}}+\textrm{H.c.}\right),
\label{e.2}
\end{eqnarray}
where $d^{\dagger}_{\mu\sigma,\mathbf{r}}$ creates a fermion of
$\mu=xz,yz$ $d$-orbital and spin
$\sigma=\uparrow,\downarrow$ at site $\mathbf{r}$. 
$t_{\parallel}$ is the NN intraorbital hopping integral in the
longitudinal direction, and $\pm t^{\prime}$ is the next NN
interorbital hopping integrals. $t_{\parallel},t^{\prime}>0$, and
$t^{\prime}$ is one order of magnitude smaller than $t_{\parallel}$. And
$t_{\parallel}$ will also be taken as the energy unit. Here
$\lambda$ is the atomic SOC strength~\cite{Kontani,Tanaka}, and it
also causes orbital polarization, but with a spin dependence.

We consider $1/N$ magnetic flux quantum per plaquette ($N$ is an
integer), i.e. $\phi=\sum_{\square}\phi_{ij}=2\pi
Ba^2/\phi_0=2\pi/N$, with $a$ the lattice constant and $\phi_0=hc/e$
the flux quantum. The Landau gauge ${\mathbf A}=(0,-Bx,0)$ and the
periodical boundary conditions (PBCs) are adopted, and the magnetic
unit cell has the size $N\times 1$. After the numerical
diagonalization of the Hamiltonian, the zero-temperature HC is
calculated through the Kubo formula~\cite{Thouless}
\begin{eqnarray}
&&{\sigma}_{\rm H}(E)={i{e^2\hbar}\over{A}}\sum_{{\cal E}_{n\mathbf{k}}<E}\sum_{{\cal E}_{m\mathbf{k}}>E}\nonumber\\
&&{{{\langle n\mathbf{k}|v_x|m\mathbf{k}\rangle\langle
m\mathbf{k}|v_y|n\mathbf{k}\rangle -\langle
n\mathbf{k}|v_y|m\mathbf{k}\rangle\langle
m\mathbf{k}|v_x|n\mathbf{k}\rangle}}\over{{({\cal
E}_{m\mathbf{k}}-{\cal E}_{n\mathbf{k}})}^2}} \label{e.3}
\end{eqnarray}
where $A=L\times L$ is the area of this 2D system, $E$ is the Fermi
energy, ${\cal E}_{n\mathbf{k}}$ is the corresponding eigenvalue of
the eigenstate $|n\mathbf{k}\rangle$ of $n$th Landau subband, and
the summation over wave vector $\mathbf{k}$ is restricted to the
magnetic Brillouin zone (MBZ): $-\pi/N\leq{k_x}a<\pi/N$ and
$-\pi\leq{k_y}a<\pi$. The velocity operator is defined as
$\mathbf{v}=(i/\hbar)[H,\mathbf{R}]$, with $\mathbf{R}$ as the
position operator of fermions. When $E$ falling in energy gaps, we
can rewrite $\sigma_{\rm H}$ as $\sigma_{\rm H}(E)=\sum_{{\cal
E}_n<E}\sigma^{(n)}_{\rm H}=e^2/h\sum_{{\cal E}_n<E}C_n$, where
$\sigma^{(n)}_{\rm H}$ and $C_n$ are the HC and the Chern
number~\cite{Thouless} of the $n$th completely filled subband,
respectively.

Now let us introduce the Berry connection {${\mathbf A}_n({\mathbf
k})=i\langle n{\mathbf k}|\nabla_{\mathbf k}|n{\mathbf k}\rangle$},
and the Berry curvature {${\mathbf \Omega}_{n}({\mathbf
k})=\nabla_{\mathbf k}\times {\mathbf A}_n({\mathbf k})$}. With these
definitions, the quantized HC of a completely filled $n$th subband
can be written as $(h/e^2)\sigma^{(n)}_{\rm
H}=(1/{2\pi})\int\int_{\rm MBZ} {\mathbf \Omega}^{z}_{n}({\mathbf
k})d^2{\mathbf k}=(1/{2\pi})\oint_{\rm MBZ}{\mathbf A}_n({\mathbf
k})\cdot d{\mathbf k}=\Gamma^{(n)}_{\rm MBZ}/2\pi$, where
{$\Gamma^{(n)}_{\rm MBZ}$} is the Berry phase of the cyclic
evolution of the $n$th eigenstate {$|n{\mathbf k}\rangle$} along the
MBZ boundary~\cite{Thouless}. While for a partially filled $n$th
subband, the non-qunatized part of the HC can also be written as
$(h/e^2)\sigma^{(n)}_{\rm H}=(1/{2\pi})\oint_{\rm FS}{\mathbf
A}_n({\mathbf k})\cdot d{\mathbf k}=\Gamma^{(n)}_{\rm FS}/2\pi$,
where {$\Gamma^{(n)}_{\rm FS}$} now is the Berry phase of the cyclic
evolution of {$|n{\mathbf k}\rangle$} along the FS~\cite{Haldane2}.
It is also instructive to introduce the sum of Berry curvatures over
the occupied subbands (for each ${\mathbf k}$): ${\mathbf
\Omega}^{z}({\mathbf k})=\sum_{{\cal E}_n<E}{\mathbf
\Omega}^{z}_{n}({\mathbf k})$.

\begin{figure}
\vspace{0.32in} \hspace{-1.1in}
\includegraphics[scale=0.5]{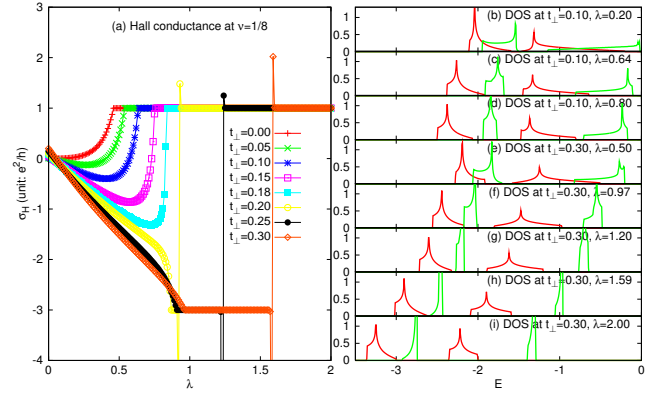}
\vspace{-0.1in} \caption{(color online). (a) Hall conductance versus
$\lambda$ (unit: $t_{\parallel}$) of $p$-band spinless fermions at
$\nu=1/8$ with $N=4$ and various $t_{\perp}$'s. (b)-(i) The DOS for
some $t_{\perp}$'s and $\lambda$'s in (a).} \label{f.1}
\end{figure}

{\it $p$-band spinless fermions. (a) Hall conductances.---}An
overall picture of the HC $\sigma_{\rm H}$ calculated by
Eq.~(\ref{e.3}) is shown in Fig.~\ref{f.1} for $p$-band spinless
fermions [Eq.~(\ref{e.1})] with $N=4$ (i.e., the flux strength
$\phi={1\over{4}}\times{2\pi}$), $2048\times2048$ lattice sites,
fermion filling $\nu=1/8$, and various $t_{\perp}$'s.

In the case of $\lambda=0$, the lowest two subbands (each
contributes ${1\over{8}}$ to $\nu$) are not separated; they give
rise to a total Chern number $+2$. With $\lambda$ increasing from
$0$ to $2.0$ one sees a systematic evolution of $\sigma_{\rm H}$
versus $\lambda$; for smaller $t_{\perp}$'s, there is a quantum
critical point (QCP) $\lambda_{c1}$ at which the lowest two subbands
begin to separate; for larger $t_{\perp}$'s, besides the first QCP
${\lambda_{c1}}$,
there is another QCP ${\lambda_{c2}}$ at which $\sigma_{\rm H}$
exhibit a quantized jump.

At smaller $t_{\perp}$'s ($t_{\perp}=0.00,\dots,0.18$), for
$\lambda>{\lambda_{c1}}$ (e.g. $\lambda_{c1}\simeq0.64$ for
$t_{\perp}=0.10$), the lowest two subbands are well separated since
$\lambda$ lifts the $p$-orbital degeneracy and induces a finite
energy gap; the lowest subband is occupied by $p_x+ip_y$ fermions
while the second lowest one by $p_x-ip_y$ fermions, each subband
carrying a Chern number $+1$, and $\sigma_{\rm H}=+1e^2/h$ (i.e.
$C_{1}=+1$) at $\nu=1/8$.

At larger $t_{\perp}$'s ($t_{\perp}=0.20,0.25,0.30$), for
$\lambda>{\lambda_{c1}}$ (e.g. $\lambda_{c1}\simeq0.97$ for
$t_{\perp}=0.30$), the lowest two subbands are also well separated;
however, the lowest subband occupied by $p_x+ip_y$ fermions now
gives a Chern number $C_{1}=-3$. When $\lambda$ increases further to
another QCP ${\lambda_{c2}}$ (e.g. $\lambda_{c2}\simeq1.59$ for
$t_{\perp}=0.30$), $\sigma_{\rm H}$ exhibit a quantized jump from
$-3e^2/h$ to $+1e^2/h$ at $\nu=1/8$ (i.e. $C_{1}$ changes from $-3$
to $+1$).

\begin{table}
\caption{\label{t.1} Topological QPTs of $p$-band spinless fermions. }
\begin{tabular}{cccc}
\hline\hline
$(N,\nu,t_{\perp})$ & $\lambda_{c1}$ & $\lambda_{c2}$ & quantized jump of $\sigma_{\rm H}$ at $\lambda_{c2}$ \\
\hline
 $(6,1/12,0.25)$  & $0.70$ & $1.16$ & $-5e^2/h\longrightarrow+1e^2/h$\\
 $(8,1/16,0.20)$  & $0.50$ & $0.89$ & $-7e^2/h\longrightarrow+1e^2/h$\\
 $(8,3/16,0.25)$  & $0.41$ & $0.59$ & $-5e^2/h\longrightarrow+3e^2/h$\\
 $(12,1/24,0.20)$ & $0.42$ & $0.92$ & $-11e^2/h\longrightarrow+1e^2/h$\\
 $(12,3/24,0.20)$ & $0.29$ & $0.45$ & $-9e^2/h\longrightarrow+3e^2/h$\\
 $(16,1/32,0.10)$ & $0.24$ & $0.44$ & $-15e^2/h\longrightarrow+1e^2/h$\\
 $(16,3/32,0.20)$ & $0.26$ & $0.53$ & $-13e^2/h\longrightarrow+3e^2/h$\\
 $(16,5/32,0.20)$ & $0.20$ & $0.26$ & $-11e^2/h\longrightarrow+5e^2/h$\\
\hline\hline
\end{tabular}
\end{table}

The above behaviors of HCs have also been verified by further
numerical calculations of the cases with $N=6-16$, various
$t_{\perp}$'s, and various $\nu$'s (Table \ref{t.1}).

\begin{figure}
\vspace{1.3in} \hspace{-0.8in}
\includegraphics[scale=0.64]{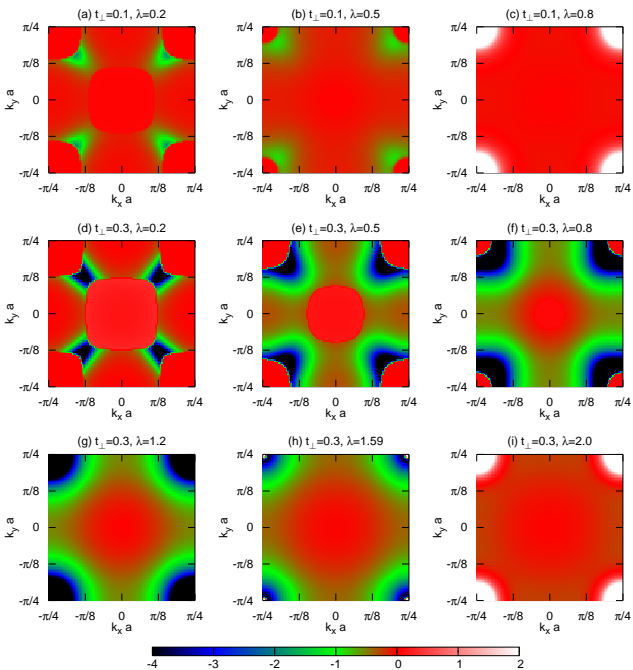}
\caption{(color online). Intensity plots of Berry curvatures
${\mathbf \Omega}^{z}({\mathbf k})$ in the RMBZ of $p$-band fermions
at $\nu=1/8$ with $N=4$, various $t_{\perp}$'s and $\lambda$'s.}
\label{f.2}
\end{figure}

{\it (b) Berry curvatures.---}In order to reveal the non-trivial
topological properties, we plot in Fig.~\ref{f.2} the Berry
curvatures ${\mathbf \Omega}^{z}({\mathbf k})$, in the reduced MBZ
(RMBZ) ($-\pi/N\leq{k_x}a,{k_y}a<\pi/N$) for some typical parameters
corresponding to Fig.~\ref{f.1}.

We first look into the case of $t_{\perp}=0.1$ which has only one
$\lambda_c$ (see Fig.~\ref{f.1}). For
$\lambda<{\lambda_{c1}}\simeq0.64$, e.g. $\lambda=0.2$
[Fig.~\ref{f.2}(a)], ${\mathbf \Omega}^{z}({\mathbf k})$ displays
the FS topology of two subbands: a hole FS of the lowest subband
near four corners of the RMBZ, and an fermion FS of the second
lowest subband near the center of the RMBZ. There are four small
negative-${\mathbf \Omega}^{z}({\mathbf k})$ regions near the hole
FS. When $\lambda$ increases, the two subbands starts to separate
and two FSs shrinks gradually [Fig.~\ref{f.2}(b)]. For
$\lambda>{\lambda_{c1}}$, e.g. $\lambda=0.8$ [Fig.~\ref{f.2}(c)],
the two subbands separates completely, both FSs vanishes, and
${\mathbf \Omega}^{z}({\mathbf k})$ displays four maxima at the four
RMBZ corners (these maxima contribute to the quantized $C_{1}=+1$).

We then analyze the case of $t_{\perp}=0.3$ which has two
$\lambda_c$'s (see Fig.~\ref{f.1}). For a small $\lambda=0.2$
[Fig.~\ref{f.2}(d)], ${\mathbf \Omega}^{z}({\mathbf k})$ also
displays the FS topology of two subbands. ${\mathbf
\Omega}^{z}({\mathbf k})$ displays four negative regions between the
hole and particle FSs. When $\lambda$ increases, e.g.
$\lambda=0.5,0.8$ [Figs.~\ref{f.2}(e) and (f)], negative-${\mathbf
\Omega}^{z}({\mathbf k})$ regions also increase, change their
topology, and enclose the four RMBZ corners, with the two FSs
shrinking. When $\lambda$ increases and passes
${\lambda_{c1}}=0.97$, e.g. $\lambda=1.2$ [Fig.~\ref{f.2}(g)], the
two subbands separates completely, and both FSs vanishes, and
${\mathbf \Omega}^{z}({\mathbf k})$ displays four negative-${\mathbf
\Omega}^{z}({\mathbf k})$ minima at the four RMBZ corners (these
minima contribute to the quantized $C_{1}=-3$). Near
${\lambda_{c2}}\simeq1.59$ [Fig.~\ref{f.2}(h)], four
negative-${\mathbf \Omega}^{z}({\mathbf k})$ minima at the RMBZ
corners begin to vanish. When $\lambda$ increases further and passes
${\lambda_{c2}}$, e.g. $\lambda=2.0$ [Fig.~\ref{f.2}(i)], the two
subbands separates completely and four negative-${\mathbf
\Omega}^{z}({\mathbf k})$ minima vanishes, and are replaced by
positive-${\mathbf \Omega}^{z}({\mathbf k})$ maxima at the RMBZ
corners (these maxima contribute to the quantized $C_{1}=+1$ again).

{\it (c) Edge states.---}An alternative way to reveal different
topological characters and QPTs is to calculate the edge
states~\cite{Hatsugai1}. Now as an illustration, we take a cylinder
of square lattice of the size $64\times\infty$ with $N=4$ (i.e., the
flux strength $\phi={1\over{4}}\times{2\pi}$), $t_{\perp}=0.30$ and
various $\lambda$'s, and apply open boundary condition (OBC) in $x$
direction and PBC in $y$ direction.

\begin{figure}
\hspace{-0.1in}
\includegraphics[scale=0.9]{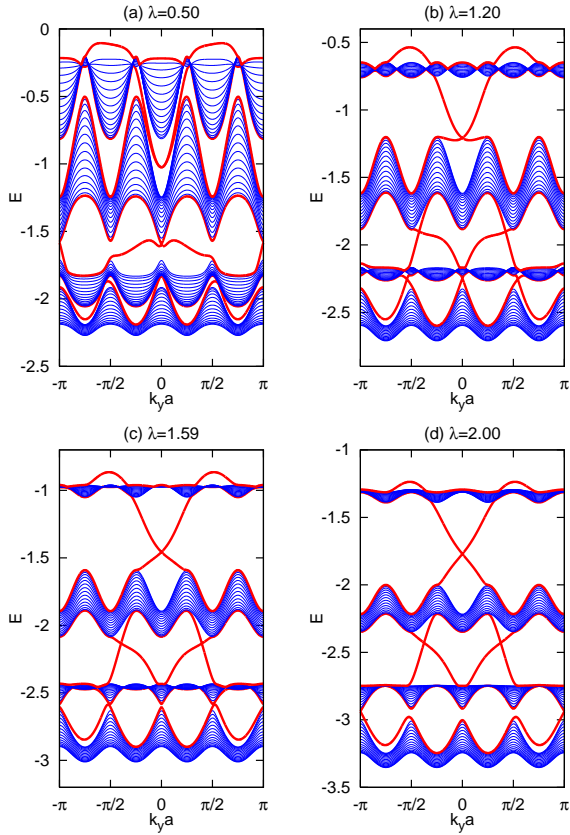}
 \vspace{-0.1in}
\caption{(color online). $E(k_y)$ of lowest four subbands and edge
states [shown as thick (red) lines] of $p$-band fermions in a
$64\times \infty$ cylinder with $t_{\perp}=0.30$ and various
$\lambda$'s.} \label{f.3}
\end{figure}

Chern numbers of the bulk subbands are intimately related to the
winding numbers of the corresponding edge states~\cite{Hatsugai1}.
We here concentrate on the edge states between the lowest two
subbands shown in Fig.~\ref{f.3}. For
$0<\lambda<\lambda_{c1}\simeq0.97$ (see the $t_{\perp}=0.30$ curve
in Fig.~\ref{f.1}), e.g. $\lambda=0.50$ [Fig.~\ref{f.3}(a)], there
is one edge state winding between the lowest two subbands, however,
the Chern number of the lowest subband $C_{1}$ is not well defined
since the energy overlap of the two subbands. For
$\lambda_{c1}<\lambda<\lambda_{c2}\simeq1.59$, e.g. $\lambda=1.20$
[Fig.~\ref{f.3}(b)], there is one edge state winding three times
from the lowest subband to the upper one then back to the lowest one
which corresponds to $C_{1}=-3$. While $\lambda>\lambda_{c2}$, e.g.
$\lambda=2.00$  [Fig.~\ref{f.3}(d)], there is another edge state
winding only once from the upper subband to the lowest one then back
to the upper one which corresponds to $C_{1}=+1$.

The continuum spectra of this cylinder also give further
descriptions about the correspondence between the quantized jumps of
the HCs and the topological evolutions of bulk
spectra~\cite{Haldane1,Hatsugai2,Wang1}. When approaching
$\lambda_{c2}$, four pairs of Dirac cones begin to form between the
lowest two subbands, and each pair of Dirac cones touch at one Dirac
point when $\lambda_{c2}\simeq1.59$  [Fig.~\ref{f.3}(c)]. Meanwhile,
a topological QPT happens at $\lambda_{c2}$, and a Chern number $+4$
is transferred from the upper subband to the lowest one, through
abrupt changes of Berry curvatures near Dirac points
[Fig.~\ref{f.2}(h)]. On contrary, at the first QCP
$\lambda_{c1}\simeq0.97$, there is no Dirac point, and thus no
quantized jumps of the HC.

\begin{figure}[!htb]
  \vspace{-0.0in}
  \hspace{-0.15in}
 \includegraphics[scale=0.59]{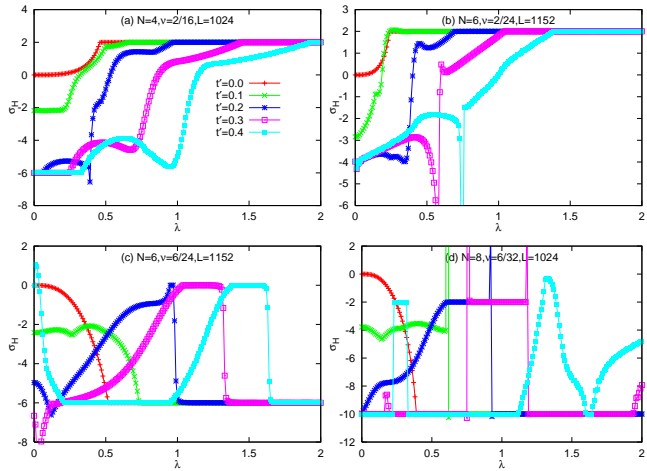}
  \vspace{-0.25in}
  \caption{(color online).  $\sigma_{\rm H}$ (unit: $e^2/h$)
versus $\lambda$ (unit: $t_{\parallel}$) of $d$-band spinful
fermions at various $N$'s, $\nu$'s and $t^{\prime}$'s.} \label{f.4}
\end{figure}

{\it $d$-band spinful fermions.---}Some typical examples of the HC
$\sigma_{\rm H}$ calculated by Eq.~(\ref{e.3}) are shown in
Fig.~\ref{f.4} for $d$-band spinful fermions [Eq.~(\ref{e.2})] with
various $N$'s, $\nu$'s and $t^{\prime}$'s. We note that TFL-to-IQH
transitions occur frequently when tuning $\lambda$ to critical
values $\lambda_c$'s. Since now we have two spin components,
$\sigma_{\rm H}$ may change either $2Ne^2/h$ [e.g. the
$t^{\prime}=0.3$ case in Fig.~\ref{f.4}(a)] or $Ne^2/h$ [e.g. the
$t^{\prime}=0.4$ case in Fig.~\ref{f.4}(c)], after passing a TFL
region or a quantized jump.

{\it Summary and discussion.---}We present that in a multi-orbital
system with intraorbital and interorbital hopping integrals, the HC
may exhibit various topological QPTs induced by on-site orbital
polarization: IQH plateau transitions, and TFL to IQH transitions.
Berry curvatures and edge states give further insights to reveal
different topological characters and QPTs. Such topological QPTs are
demonstrated in both a $p$-band spinless fermionic system in optical
lattice, and a $d$-band spinful fermionic system related to giant
orbital Hall effects in transition metals and their compounds. In
optical lattices, artificial magnetic fields can be created by laser
assisted tunneling between internal atomic states, or time-varying
quadrupole potential~\cite{Demler}.

This work was supported by the National Natural Science Foundation
of China (No. 10904130), the State Key Program for Basic Researches
of China (No. 2006CB921802), and the Start-Up Funding at ZJNU.


\begin{thebibliography}{99}
\bibitem{Isacsson} A. Isacsson and S. M. Girvin, Phys. Rev. A {\bf 72}, 053604
(2005); W. V. Liu and C. Wu, {\it ibid.} {\bf 74}, 013607 (2006); A.
B. Kuklov, Phys. Rev. Lett. {\bf 97}, 110405 (2006); C. Wu, W. V.
Liu, J. E. Moore, and S. Das Sarma, {\it ibid.} {\bf 97}, 190406
(2006).




\bibitem{Kohl} M. K\"{o}hl {\it et al.},
Phys. Rev. Lett. {\bf 94}, 080403 (2005).


\bibitem{Muller} T. M\"{u}ller, S. F\"{o}lling, A. Widera, and I. Bloch, Phys. Rev. Lett. {\bf 99},
200405 (2007).

\bibitem{Anderlini} M. Anderlini {\it et al.}, Nature (London) {\bf
448}, 452 (2007). 



\bibitem{Wu2} C. Wu, D. Bergman, L. Balents, and S. Das Sarma, Phys. Rev. Lett.
{\bf 99}, 070401 (2007).

\bibitem{Wu3} C. Wu, Phys. Rev. Lett. {\bf 100}, 200406 (2008).

\bibitem{Zhao} E. Zhao and W. V. Liu, Phys. Rev. Lett. {\bf 100}, 160403 (2008).


\bibitem{Haldane1} F. D. M. Haldane, Phys. Rev. Lett. {\bf 61}, 2015 (1988).

\bibitem{Wu4} C. Wu, Phys. Rev. Lett. {\bf 101}, 186807 (2008).




\bibitem{Valenzuela} S. O. Valenzuela and
M. Tinkham, Nature (London) {\bf 442}, 176 (2006).

\bibitem{Stern} N. P. Stern {\it et al.}, Phys. Rev. Lett. {\bf 97}, 126603 (2006). 

\bibitem{Kimura} T. Kimura {\it et al.}, Phys.
Rev. Lett. {\bf 98}, 156601 (2007). 




\bibitem{Kontani} H. Kontani {\it et al.}, Phys. Rev. Lett. {\bf 100}, 096601 (2008); {\it ibid.} {\bf
102}, 016601 (2009). 

\bibitem{Tanaka} T. Tanaka {\it et al.}, Phys. Rev. B {\bf 77}, 165117 (2008); T. Tanaka and H.
Kontani, {\it ibid.} {\bf 77}, 195129 (2008).  


\bibitem{Guo} G. Y. Guo, S. Murakami, T. W. Chen, and N. Nagaosa, Phys. Rev. Lett. {\bf 100}, 096401 (2008).


\bibitem{Thouless} D. J. Thouless, M. Kohmoto, M. P. Nightingale, and M. den Nijs, Phys. Rev. Lett. {\bf 49}, 405
(1982).

\bibitem{Haldane2} F. D. M. Haldane, Phys. Rev. Lett. {\bf 93}, 206602 (2004).


\bibitem{Umucalilar} R. O. Umucal{\i}lar and M. \"{O}. Oktel, Phys. Rev. A {\bf 78}, 033602
(2008).


\bibitem{Hatsugai1} Y. Hatsugai, Phys. Rev. Lett. {\bf 71}, 3697 (1993).


\bibitem{Hatsugai2} Y. Hatsugai and M. Kohmoto, Phys. Rev. B {\bf 42}, 8282 (1990).

\bibitem{Wang1} Y. F. Wang and C. D. Gong, Phys. Rev. Lett. {\bf 98}, 096802 (2007);
Y. F. Wang, Y. Zhao, and C. D. Gong, Phys. Rev. B {\bf 78}, 045301 (2008).

\bibitem{Demler} D. Jaksch and P. Zoller, New J. Phys. {\bf 5}, 56 (2003);
E. J. Mueller, Phys. Rev. A {\bf 70}, 041603(R) (2004); A.S.
S{\o}rensen, E. Demler, and M. D. Lukin, Phys. Rev. Lett. {\bf 94},
086803 (2005); K. Osterloh {\it et al.}, {\it ibid.} {\bf 95},
010403 (2005); Y. J. Lin, {\it et al.}, {\it ibid.} {\bf 102},
130401 (2009)

\end{thebibliography}
\end{document}